\documentclass[aps,prb,twocolumn,showpacs]{revtex4-1}
\usepackage{graphicx}
\usepackage{dcolumn}
\usepackage{bm}
\usepackage{color}
\usepackage{amsmath}
\usepackage[export]{adjustbox}

\newcommand{\ket}[1]{\left|#1\right\rangle}

\renewcommand{\Im}{{\rm Im}}

\newcommand{\Tr}{{\rm Tr}}

\newcommand{\tip}{{\rm T}}

\newcommand{\sub}{{\rm S}}
\newcommand{\cen}{{\rm C}}
\newcommand{\dev}{{\rm D}}

\newcommand{\su}{{\uparrow}}
\newcommand{\sd}{{\downarrow}}
\newcommand{\didv}{{{\rm d}{I}/{\rm d}{V}}}
\newcommand{\rmd}{{\rm d}}

\begin{document}

\title[Simulation of inelastic spin flip excitations and Kondo effect in STM spectroscopy]
      {Simulation of inelastic spin flip excitations and Kondo effect in STM spectroscopy of magnetic molecules on metal substrates}

\author{David Jacob}
\affiliation{Nano-Bio Spectroscopy Group, Dpto. de F\'isica de Materiales,
  Universidad del Pa\'is Vasco UPV/EHU, Avenida Tolosa 72, E-20018 San
  Sebasti\'an, Spain}
\affiliation{IKERBASQUE, Basque Foundation for Science, Mar\'ia D\'iaz de Haro 3,
  E-48013 Bilbao, Spain}

\email{david.jacob@ehu.eus}

\begin{abstract}
  Single-ion magnetic anisotropy in molecular magnets leads to spin flip excitations that can 
  be measured by inelastic scanning tunneling microscope (STM) spectroscopy. Here I present a semi ab initio
  scheme to compute the spectral features associated with inelastic spin flip excitations and Kondo effect
  of single molecular magnets. To this end density functional theory calculations
  of the molecule on the substrate are combined with more sophisticated many-body techniques
  for solving the Anderson impurity problem of the spin-carrying orbitals of the magnetic molecule
  coupled to the rest of the system, containing a phenomenological magnetic 
  anisotropy term.
  For calculating the STM spectra an exact expression for the $\didv$ in the ideal STM limit,
  when the coupling to the STM tip becomes negligibly small, is derived.
  In this limit the $\didv$ is simply related to the spectral function of the molecule-substrate system.
  For the case of an Fe porphyrin molecule on the Au(111) substrate, the
  calculated STM spectra are in good agreement with recently measured STM spectra,
  showing the typical step features at finite bias associated with spin flip excitation
  of a spin-1 quantum magnet.
  For the case of Kondo effect in Mn porphyrin on Au(111), the agreement with the experimental spectra
  is not as good due to the neglect of quantum interference in the tunneling.
\end{abstract}

\maketitle

\section{Introduction}
\label{sec:Intro}

The electronic and magnetic properties of nanoscale quantum magnets, such as magnetic atoms,
clusters and molecules on conducting substrates, can be strongly affected by their environment,
and can be investigated by means of scanning tunneling microscopy (STM)~\cite{Binning:PRL:1982}.
Specifically, STM spectroscopy (STS) allows to measure the electronic and magnetic excitations
of nanoscale systems~\cite{Strocio:book:1993,Chen:book:1993,Wiesendanger:book:1994,Hamers:chapter:2001}.
For example, step features at finite bias voltages in the differential conductance ($\didv$) spectra
measured by STS can be related to inelastic spin flip excitations, associated with the magnetic
anisotropy of the nanoscale magnet~\cite{Hirjibehedin:Science:2007,Zitko:PRB:2008,Fernandez-Rossier:PRL:2009,Lorente:PRL:2009}.
On the other hand, the appearance of a zero-bias anomaly in the STM spectrum points to a Kondo effect
(see e.g. the book by Hewson~\cite{Hewson:book:1997}) due to a degenerate magnetic ground state screened
by the conduction electrons~\cite{Madhavan:Science:1998,Schiller:PRB:2000,Ujsaghy:PRL:2000}.
Both phenomena have also been observed simultaneously, for example in the case of Co on CuN
substrates, where the magnetic anisotropy leads to a splitting of the spin-3/2 ground state
into two doublet states~\cite{Otte:NPhys:2008,Oberg:NNano:2014}.

Moreover, interesting information about the system is also encoded in the actual lineshapes
of the spectral features associated with both phenomena in the measured $\didv$ spectra.
For example, a Kondo resonance gives rise to a zero-bias anomaly that
is generally well described by a Fano lineshape~\cite{Fano:PR:1961,Schiller:PRB:2000,Ujsaghy:PRL:2000}
or a generalized Frota lineshape~\cite{Frota:PRB:1992,Karan:PRL:2015}. The actual shape of this Fano/Frota feature
reveals information about the spin-carrying orbitals involved in the Kondo effect~\cite{Frank:PRB:2015},
or the voltage drop within a molecular junction~\cite{Karan:PRL:2015}.
Similarly, the actual lineshapes of the spin flip excitation steps in STS of magnetic porphyrin molecules
on metal substrates contains information about the orbitals involved in the spin flip excitations~\cite{Rubio-Verdu:NCommP:2018}.

On the theory side, most of the phenomenology of STM spectroscopy of nanoscale quantum magnets on
conducting substrates can be described in terms of Kondo type and Anderson type impurity
models, which capture the Kondo effect as well as spin flip excitations by inclusion of
a magnetic anisotropy term into the model~\cite{Zitko:PRB:2008,Fernandez-Rossier:PRL:2009,Lorente:PRL:2009,Zitko:NJP:2010,Hurley:PRB:2011,Oberg:NNano:2014,Delgado:SS:2014,Ternes:NJP:2015,Jacob:EPJB:2016,Jacob:PRB:2018}.
In combination with experiments these model Hamiltonian 
calculations have revealed interesting effects such as the renormalization of single-ion magnetic
anisotropy by Kondo exchange coupling to the conduction electrons~\cite{Oberg:NNano:2014,Hiraoka:NComm:2017}. 
Ab initio density functional theory (DFT) calculations, on the other hand, yield valuable insights
about the molecular orbitals hosting the spin and the charging state of the molecule~\cite{Karan:NL:2018,Rubio-Verdu:NCommP:2018},
but cannot account for the dynamic correlations that give rise to Kondo effect and spin flip excitations. 

By combining ab initio DFT calculations with impurity model calculations, it is possible
to gain further insights into the often rather complex situation encountered in real nanoscale
systems~\cite{Jacob:PRL:2009,Surer:PRB:2012,Jacob:PRB:2013,Jacob:JPCM:2015,Minamitani:PRB:2015,Dang:PRB:2016,Droghetti:PRB:2017}.
For example, DFT plus impurity solver calculations attribute the Kondo resonances
in the STS of manganese phthalacyanine (MnPc) on the lead substrate and of manganese porphyrin
on gold to underscreened Kondo effects in the Mn $z^2$-orbital, strongly influenced by charge
fluctuations~\cite{Jacob:PRB:2013,Karan:PRL:2015}. Also the Fano lineshape in the STS measured
for the Co on Cu(001) could be attributed in this way to a Kondo peak in the Co $z^2$-orbital~\cite{Jacob:JPCM:2015,Frank:PRB:2015}. 

In these cases the low-bias transport characteristics can be straight forwardly calculated
in the phase coherent approximation from the correlated transmission function~\cite{Jacob:PRL:2009,Jacob:JPCM:2015,Droghetti:PRB:2017},
which takes into account quantum interference effects as well as elastic many-body effects such as the Kondo effect,
leading e.g. to Fano behavior in the $\didv$. On the other hand, inelastic electron scattering resulting
from electron-electron interactions is neglected in this approach~\cite{Droghetti:PRB:2017}.
While in the case of the Kondo effect, the inelastic corrections to the low-bias transport properties
are actually very small~\cite{Choi:JCP:2017}, this is naturally not the case for the inelastic spin flip
excitations, for which inelastic many-body scattering plays of course a crucial role.
In principle one would thus have to make use of the Meir-Wingreen equation for computing
the current via a nanoscopic region~\cite{Meir:PRL:1992}. This requires to
solve the many-body problem out of equilibrium which is in general a difficult and
computationally very demanding task.

Instead here we make use of an exact limit of the Meir-Wingreen equation when the coupling to the STM tip
becomes very weak compared to the coupling to the substrate and all the applied voltage drops at
the STM tip~\cite{Jacob:NL:2018}. In this \emph{ideal STM limit} the molecule is in quasi equilibrium with the
substrate, and the $\didv$ can be expressed simply in terms of the \emph{equilibrium} spectral
function of the molecule. For finite but small coupling to the STM tip the ideal STM limit becomes
an approximation, that captures elastic (e.g. Kondo effect) as well a inelastic (e.g. spin flip excitations)
many-body effects encoded in the spectral function, but neglects quantum interference phenomena
in the tunneling from the tip to the molecule.

The paper is organized as follows: In Sec.~\ref{sec:Method} the methodology for computing the STM spectra
of magnetic molecules on conducting substrates is introduced. A special focus lies on the derivation of the
\emph{ideal STM limit} (Sec.~\ref{ssec:STM_limit}). In Sec.~\ref{sec:Results} the methodology is applied
to the calculation of the electronic structure and STM spectra of Fe and Mn porphyrin molecules on the Au(111)
substrate. Similar systems only differing by the type of ligands have been recently measured in a number of
experiments~\cite{Karan:PRL:2015,Karan:NL:2018,Li:SciAdv:2018,Rubio-Verdu:NCommP:2018}.
In Sec.~\ref{sec:Conclusions} the paper concludes with a couple of general remarks regarding the results
and methodology.

\section{Methodology}
\label{sec:Method}

\subsection{Setup}
\label{ssec:Setup}

\begin{figure}
  \begin{tabular}{cc}
    \includegraphics[width=0.45\linewidth]{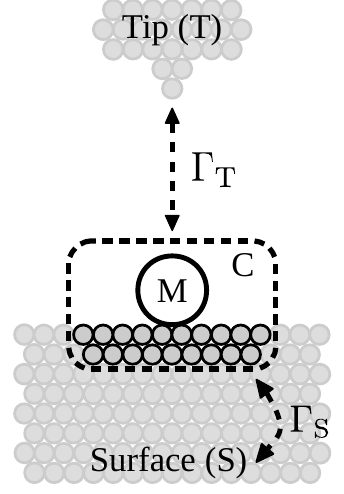} &
    \includegraphics[width=0.45\linewidth]{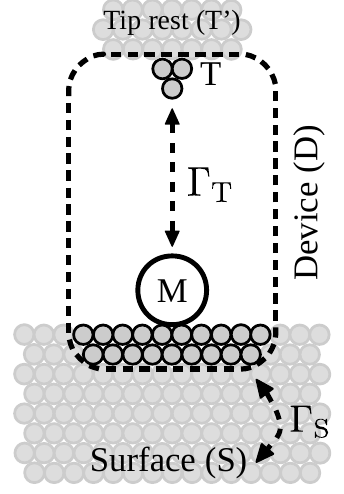}
  \end{tabular}
  \caption{\label{fig:setup}
    Left: Schematic drawing of typical STM setup for probing a molecule.
    The STM tip (T) probes the central region (C) consisting of the molecule (M)
    and part of the substrate.
    The coupling $\bm\Gamma_\tip$ of C to the STM tip T is usually much weaker
    than the coupling $\bm\Gamma_\sub$ to the rest of the substrate (S),
    $\bm\Gamma_\tip\ll\bm\Gamma_\sub$.
    Right: The DFT calculations are done for the bigger device region (D) which
    includes a small portion of the STM tip T in addition to the central region C.
    This setup allows to determine the coupling matrix $\bm\Gamma_\tip$ from first
    principles (see text for further explanations).
  }
\end{figure}

We consider the situation schematically depicted in Fig.~\ref{fig:setup}: 
A molecule (M) deposited on a metal substrate (S) is coupled weakly to a
second electrode, the STM tip (T), which serves as a probe. The T electrode
couples also weakly to part of the surface in the proximity of M. We thus
define a central region (C) which contains M and the part of the surface
coupled to T. We assume that electron-electron interactions
only take place in the central region, while the electrons in the two electrodes
T and S are (effectively) non-interacting.

The Hamiltonian of the central region coupled to the two electrodes S and T
is thus given by
\begin{equation}
  \hat{\mathcal{H}} = \hat{\mathcal{H}}_\cen + \hat{\mathcal{H}}_\sub + \hat{\mathcal{V}}_\sub + \hat{\mathcal{H}}_\tip + \hat{\mathcal{V}}_\tip 
\end{equation}
where $\hat{\mathcal{H}}_\cen$ is the Hamiltonian of the central region,
which comprises a general electron-electron interaction term:
\begin{equation}
  \label{eq:H_C}
  \hat{\mathcal{H}}_\cen = \sum_{i,j,\sigma} (\bm{H}^0_\cen)_{ij} \, \hat{d}^\dagger_{i\sigma} \hat{d}_{j\sigma}
  + \frac{1}{2} \sum_{{i,j,k,l}\atop{\sigma,\sigma^\prime}} V_{ijkl} \,
  \hat{d}^\dagger_{i\sigma} \hat{d}^\dagger_{j\sigma^\prime} \hat{d}_{l\sigma^\prime}
  \hat{d}_{k\sigma}
\end{equation}
where $\bm{H}^0_\cen$ is the one-body part of the Hamiltonian
(in matrix notation) and $V_{ijkl}$ are the matrix elements of the Coulomb interaction.

Electrode $\alpha$ ($\alpha \in \{\tip,\sub\}$),
is described by the Hamiltonian $\hat{\mathcal{H}}_\alpha=\sum_{ij\sigma} ((\bm{H}_\alpha)_{ij} +\mu_\alpha\delta_{ij}) \, \hat{c}_{\alpha,i\sigma}^\dagger \hat{c}_{\alpha,j\sigma}$
where we have also included the chemical potential $\mu_\alpha$ applied to electrode $\alpha$,
which describes the electrostatic shift of the electrode band structure induced by the chemical potential.
The coupling between C and electrode $\alpha$ is described by $\hat{\mathcal{V}}_{\alpha} = \sum_{i,k,\sigma} (\bm{V}_\alpha)_{ki} \, \hat{c}^\dagger_{\alpha,k\sigma} \hat{d}_{i\sigma} + {\rm h.c.}$

The retarded Green's function of the central region can now be written as
\begin{equation}
  \bm{G}_\cen(\omega) = \left( \omega -\bm{H}^0_\cen -\bm\Sigma_\cen(\omega) -\bm\Sigma_\tip(\omega) -\bm\Sigma_\sub(\omega) 
  \right)^{-1}
\end{equation}
where $\bm\Sigma_\cen(\omega)$ is the many-body self-energy describing the effect of
electron-electron interactions within region C. $\bm\Sigma_\alpha(\omega)$ for $\alpha \in \{\tip,\sub\}$
on the other hand are the so-called embedding self-energies given by
\begin{equation}
  \label{eq:SE_embed}
  \bm{\Sigma}_\alpha(\omega) = \bm{V}_\alpha^\dagger \frac{1}{\omega-\mu_\alpha-\bm{H}_\alpha} \bm{V}_\alpha
\end{equation}
which describe the coupling of C to the conduction electron bath in the two electrodes.
The anti-hermitian part of the embedding self-energies yields the so-called
coupling matrices, $\bm\Gamma_\alpha=i(\bm\Sigma_\alpha^\dagger-\bm\Sigma_\alpha)$,
which describe the broadening of the central region due to the coupling to the
electrodes.

\subsection{The ideal STM limit}
\label{ssec:STM_limit}

By applying a bias voltage $eV=\mu_\tip-\mu_\sub$ between the STM
tip and the substrate a current $I$ is driven through the molecular junction.
In the ideal limit of very weak coupling to the STM tip ($\Gamma_\tip\ll\Gamma_\sub$)
and for the applied voltage dropping entirely at the STM tip ($\mu_\tip=eV$ and $\mu_\sub=0$),
the differential conductance $\didv$ can be directly related to the
\emph{equilibrium} many-body spectral function of C, $\bm{A}_\cen(\omega)=-2\,\Im\,\bm{G}_\cen(\omega)$~\cite{Jacob:NL:2018}.
The key observation is that in the \emph{ideal STM limit}, $\Gamma_\tip\rightarrow0$, the non-equilibrium GFs
of the central region become independent of the applied bias $V$, i.e. the C region is essentially
in equilibrium with the substrate. Hence the lesser GF of C reduces to its equilibrium value,  
$i\bm{G}_\cen^<(\omega)\rightarrow f_\sub(\omega)\bm{A}_\cen(\omega)$ for $\Gamma_\tip\rightarrow0$,
where $f_\sub(\omega)=f(\omega)$ is the Fermi function for the substrate, which does not depend on
the bias.

The Fermi function of the STM tip, on the other hand, is given by $f_\tip(\omega)=f(\omega-eV)$.
In the ideal STM limit, the tunneling current from the tip electrode to the central region given
by the Meir-Wingreen expression~\cite{Meir:PRL:1992}
\begin{equation}
  \label{eq:Meir-Wingreen}
  I = \frac{2e}{h} \int{\rm d}\omega \Tr\left\{ \bm\Gamma_\tip(\omega) \left[ f_\tip(\omega)\,\bm{A}_\cen(\omega) + i \bm{G}_\cen^<(\omega)\right] \right\}
\end{equation}
thus reduces to
\begin{equation}
  I  \xrightarrow[\Gamma_\tip\rightarrow0]{} \frac{2e}{h} \int{\rm d}\omega
  \left[f_\tip(\omega)+f_\sub(\omega)\right] \Tr\left\{\bm\Gamma_\tip(\omega) \bm{A}_\cen(\omega) \right\}
\end{equation}
Taking the derivative w.r.t. the bias $V$ then yields
\begin{equation}
  \frac{{\rm d}{I}}{{\rm d}{V}} \xrightarrow[\Gamma_\tip\rightarrow0]{}  \frac{2e^2}{h} \int\rmd\omega [-f^\prime(\omega-eV)] \Tr\left\{\bm\Gamma_\tip(\omega) \bm{A}_\cen(\omega) \right\}
\end{equation}
Finally, in the zero temperature limit, $-f^\prime(\omega)\rightarrow\delta(\omega)$,
and hence in the ideal STM limit ($\Gamma_\tip,T\rightarrow0$) at zero temperature, we
obtain the following expression that relates the differential conductance directly
to the equilibrium many-body spectral function of the central region:
\begin{equation}
  \label{eq:stm_didv}
  \frac{\rmd{I}}{\rmd{V}} = \frac{2e^2}{h} \Tr\left\{\bm\Gamma_\tip^0\bm{A}_\cen(eV) \right\} 
\end{equation}
where $\bm\Gamma_\tip^0$ is the \emph{equilibrium} tip coupling matrix (i.e. for $\mu_\tip=0$), evaluated at
the Fermi level ($\omega=0$), since $\bm\Gamma_\tip(eV)= -2\Im\bm{V}_\tip^\dagger (eV-eV-\bm{H}_\tip)^{-1}\bm{V}_\tip\equiv\bm\Gamma_\tip^0$
due to the electrostatic shift of the tip band structure by $\mu_\tip=eV$.

Eq.~\ref{eq:stm_didv} is essentially a generalization of the Tersoff-Hamann result for the description of
electron tunneling in STM experiments~\cite{Tersoff:PRB:1985} to the case of orbital-dependent coupling to the STM tip
(i.e. non-constant tunneling matrix elements).
Also note that (\ref{eq:stm_didv}) represents an exact limit of the Meir-Wingreen equation, fully taking into account interaction effects
in the central region, encoded in the many-body spectral function $\bm{A}_\cen(\omega)$.
Of course, for a realistic STM setup $\bm\Gamma_\tip$ is finite, and hence (\ref{eq:stm_didv}) is actually an approximation,
although a very good one, as the tip coupling $\bm\Gamma_\tip$ is usually orders of magnitude smaller
than the coupling to the substrate $\bm\Gamma_\sub$ (see also below) due to the exponential dependence
of the tunneling matrix elements on the distance to the sample.

\subsection{DFT+Anderson impurity solver calculations}
\label{ssec:DFT+AIM}

In order to calculate the $\rmd{I}/\rmd{V}$ we have thus to compute the spectral
function of the central region, taking into account the coupling to the substrate and
interaction effects in C. Here we make use of the NanoDMFT approach which combines
mean-field like Kohn-Sham (KS) DFT calculations for the interacting region C with
a more sophisticated many-body treatment for a small portion of the system
using an Anderson impurity solver technique, in order to take into account dynamic
correlation effects arising from strong electronic interactions, e.g. within the
$3d$-shell of a transition metal atom~\cite{Jacob:PRL:2009,Jacob:JPCM:2015}. 
In the case of several Anderson impurities in the central region, Dynamical Mean-Field Theory
(DMFT) adapted to nanoscale systems (NanoDMFT) can be employed~\cite{Jacob:PRB:2010a}.

As we also have to determine the coupling $\bm\Gamma_\tip$ to the STM tip, it is convenient
to include part of the STM tip into the DFT calculation, even though the coupling
of the central region to the STM tip is weak, and thus does not really influence its electronic
structure.
Hence we extend the central region and include a small part of the tip electrode T into the larger
device region (D), as depicted schematically in the right panel of Fig.~\ref{fig:setup}.
In a first step, the electronic structure of the device region D is then calculated on the DFT level,
taking into account the coupling to the substrate S and the rest of the tip electrode T' using
the ANT.G package~\cite{ANTG} in connection with the Gaussian09 quantum chemistry code~\cite{G09},
as described in more detail in previous in work~\cite{Jacob:JCP:2011}. This yields the KS GF of the device region
\begin{equation}
  \bm{G}^s_\dev(\omega) = \left( \omega -\bm{H}^0_\dev-\bm{V}_{\rm Hxc} -\bm\Sigma_{\tip^\prime}(\omega) -\bm\Sigma_\sub(\omega) 
  \right)^{-1}
\end{equation}
where $\bm{H}^0_\dev$ is the one-body part of the Hamiltonian of the device region and
$\bm{V}_{\rm Hxc}$ is the Hartree exchange-correlation (Hxc) potential which yields an effective
mean-field description of the interactions within D.
$\bm\Sigma_{\tip^\prime}$ is the embedding self-energy describing the coupling of a small part of
the STM tip T included in D to the rest of the STM tip T', and $\bm\Sigma_\sub$ the embedding
self-energy describing the coupling of C to the rest of the surface S, as before.
The self-energy $\bm\Sigma_\tip$ and corresponding coupling matrix $\bm\Gamma_\tip$
%%describing the coupling of the STM tip T to the central region C,
can now be calculated from the projection of the KS device GF onto the tip atoms
included in D,  $\bm{G}^s_\tip(\omega)=\bm{P}_\tip \,\bm{G}^s_\dev(\omega)\,\bm{P}_\tip$, as
\begin{equation}
  \label{eq:Sigma_tip}
  \bm{\Sigma}_\tip(\omega) = \bm{V}_\tip^\dagger \bm{G}^s_\tip(\omega) \bm{V}_\tip
\end{equation}
where the hopping matrix $\bm{V}_\tip$ is obtained from the off-diagonal projection of the KS Hamiltonian,
$\bm{H}^s_\dev=\bm{H_\dev^0}+\bm{V}_{\rm Hxc}$ on the T and C parts of the device region: $\bm{V}_\tip=\bm{P}_\tip\,\bm{H}^s_\dev\,\bm{P}_\cen$.

The static mean-field picture of the KS DFT calculation does not account for so-called dynamic correlation effects
originating from strong electronic interactions e.g. in the open $d$-shells of transition metal (TM) atoms, or the
$f$-shells of Lanthanide or Actinide atoms. 
Dynamic correlations give rise for example to the Kondo effect and are important for the description of spin flip
excitations of magnetic atoms and molecules on conducting substrates~\cite{Oberg:NNano:2014,Jacob:EPJB:2016,Jacob:PRB:2018}.
In the next step, we thus perform a projection onto the strongly correlated subspace, which in our case is given by
the $3d$-shell of a transition metal (TM) atom at the center of the molecule. This yields an Anderson impurity model (AIM)
describing the $3d$-shell coupled to the substrate and to the rest of the molecule. In order to take into account
dynamic correlations beyond the mean-field description, the $3d$-shell is augmented by an (effective) Coulomb interaction
term. The Hamiltonian of the $3d$-shell thus reads:
\begin{eqnarray}
  \label{eq:H_imp}
  \mathcal{H}_{3d} &=& \sum_{i,j,\sigma} h^{3d}_{ij} d_{i\sigma}^\dagger d_{j\sigma}
  + \frac{1}{2} \sum_{{i,j,k,l}\atop{\sigma,\sigma^\prime}} U_{ijkl} \,
  d^\dagger_{i\sigma} d^\dagger_{j\sigma^\prime} d_{l\sigma^\prime} d_{k\sigma} \nonumber\\
  & & + D (\hat{S}^{3d}_z)^2
\end{eqnarray}
where the indices $i,j,k,l$ now run over the impurity shell. The one-body part $h^{3d}_{ij}$ is given by the
KS Hamiltonian $\bm{H}^s_\dev$ projected onto the $3d$-shell, corrected by a double-counting (DC) term: $h^{3d}_{ij} = (H^s_\dev)_{ij}-V^{\rm dc}_{ij}$.
The latter accounts for the fact that Coulomb interaction within the $3d$-shell has already been taken into account
in the KS Hamiltonian in an effective mean-field way.
Unfortunately, the DC term $\hat{V}^{\rm dc}$ is not exactly known for DFT, and several approximation schemes are used in practice~\cite{Karolak:JESRP:2010}.
Here the so-called atomic limit or fully localized limit (FLL) is employed~\cite{Czyzyk:PRB:1994}.

Also note that the effective Coulomb interaction $U_{ijkl}$ is different from the bare one $V_{ijkl}$ due to
screening by electron-hole pairs in the rest of the system.
It is in principle possible to calculate the screened Coulomb interaction $U_{ijkl}$~\cite{Jacob:JPCM:2015} for example within the
constrained random phase approximation (cRPA)~\cite{Aryasetiawan:PRB:2006}.
Instead here we make a reasonable guess for the Coulomb matrix elements which leads to good agreement between the calculated
STM spectra and the experimental ones (see below).
Finally, the last term in (\ref{eq:H_imp}) yields a phenomenological description of the single-ion magnetic anisotropy (MA)
arising from crystal-field splitting of the impurity levels in combination with spin-orbit coupling. As was shown in previous
work~\cite{Oberg:NNano:2014,Jacob:EPJB:2016,Jacob:PRB:2018} the MA term gives rise to inelastic spin-flip excitations in the
spectral function of the impurity shell. 

The coupling between the impurity shell and the rest of the system (i.e. the bath) consisting of the the substrate and the rest of the molecule
is described by the embedding self-energy of the $3d$-shell, usually called hybridization function (matrix) 
and denoted by $\bm\Delta(\omega)$ in the context of the AIM.
The hybridization function matrix of the $3d$-shell can be obtained by reverse engineering from the projection of the
device GF onto the $3d$-subspace, $\bm{G}_{3d} = \bm{P}_{3d}\,\bm{G}_\dev\,\bm{P}_{3d}$, as
\begin{equation}
  \label{eq:hybfunc}
  \bm{\Delta}_{3d}(\omega) = \omega - \bm{H}^s_{3d} - [\bm{G}_{3d}(\omega)]^{-1}
\end{equation}
where $\bm{H}^s_{3d}$ is the KS Hamiltonian projected onto the $3d$-shell. 

The impurity Hamiltonian (\ref{eq:H_imp}) together with the hybridization function (\ref{eq:hybfunc})
completely defines the AIM.
Here as in previous works~\cite{Jacob:PRL:2009,Karolak:PRL:2011,Jacob:PRB:2013,Jacob:JPCM:2015}
we make use of the so-called one-crossing approximation (OCA) for solving the AIM~\cite{Pruschke:ZPB:1989,Haule:PRB:2001,Haule:PRB:2010}.
OCA consists in a diagrammatic expansion of the propagators $G_n(\omega)$ associated with the
many-body eigenstates $\ket{n}$ of the \emph{isolated} impurity Hamiltonian (\ref{eq:H_imp}) in terms of
the hybridization function $\bm\Delta_{3d}(\omega)$, summing only a subset of diagrams
(only those where conduction electron lines cross at most once) to infinite order. 
Once the impurity problem is solved, the many-body spectral function of the impurity shell $\bm{A}_{3d}(\omega)$
and the many-body self-energy $\bm\Sigma_{3d}(\omega)$ describing the interaction effects within the $3d$-shell
are obtained. 
The \emph{correlated} GF of the central region C is then given by
\begin{equation}
  \bm{G}_\cen(\omega) = \left( \omega -\bm{H}^s_\cen - \bm\Sigma_\cen^{\rm MB}(\omega) -\bm\Sigma_\sub(\omega) 
  \right)^{-1}
\end{equation}
where $\bm\Sigma_\cen^{\rm MB}(\omega)$ is the many-body self-energy, describing interaction effects
within the C region. $\bm\Sigma_\cen^{\rm MB}(\omega)$ comprises the static mean-field like
Hxc term $\bm{V}_{\rm Hxc}$ of the DFT calculation and the dynamic
many-body correction $\bm\Sigma_{3d}(\omega)-\bm{V}_{\rm dc}$ for the correlated $3d$-subspace:
$\bm\Sigma_\cen(\omega)=\bm{V}_{\rm Hxc}+\bm\Sigma_{3d}(\omega)-\bm{V}_{\rm dc}$.
From the GF $\bm{G}_\cen(\omega)$, the spectral function $\bm{A}_\cen(\omega)=-\Im\,\bm{G}_\cen(\omega)$,
and thus the STM spectrum of the molecule, taking into account dynamic correlation effects in the
$3d$-shell of the TM center, can be calculated via (\ref{eq:stm_didv}).

\section{Results}
\label{sec:Results}

\subsection{Spin flip excitations of FeP on Au(111)}

We now apply the above developed methodology to the description of spin excitations measured
recently by STS of an Fe tetraphenylporphyrin sulfonate (FeTTPS) molecule on the Au(111)
surface~\cite{Karan:NL:2018}. Similar Fe porphyrin type molecules have recently been
measured by other groups~\cite{Li:SciAdv:2018,Rubio-Verdu:NCommP:2018}.

\begin{figure}
  \begin{tabular}{cc}
    \includegraphics[width=0.45\linewidth]{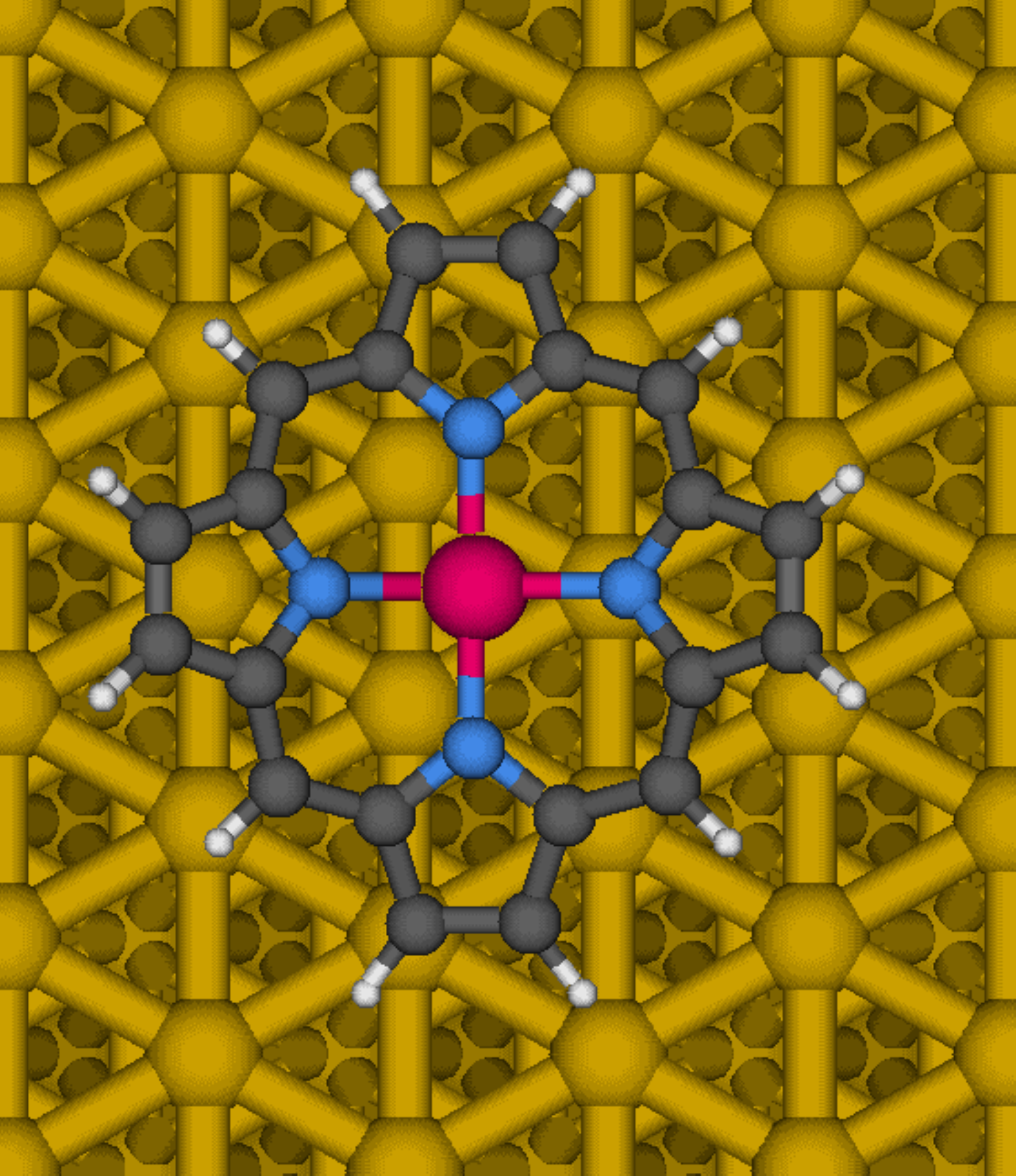} &
    \includegraphics[width=0.45\linewidth]{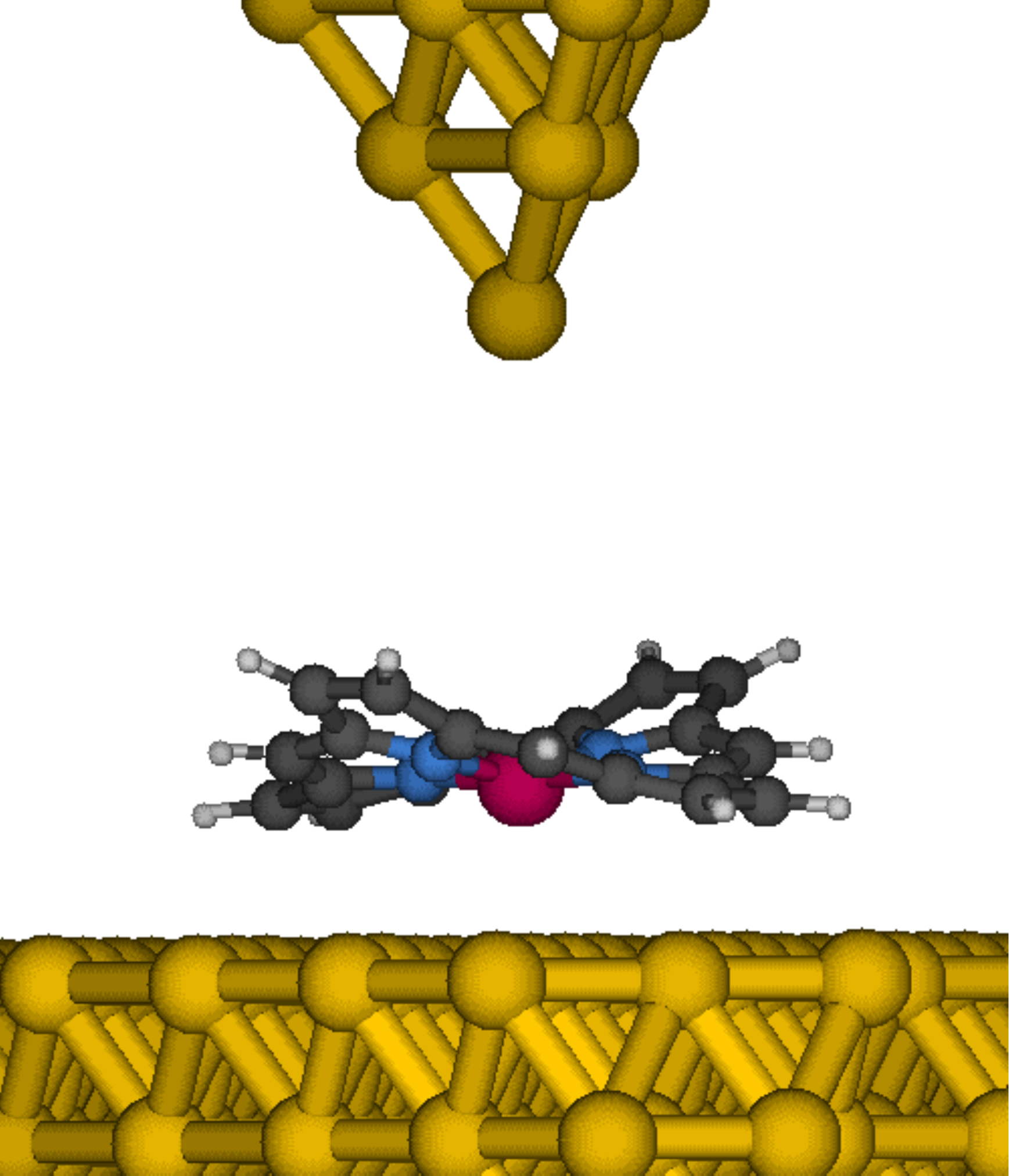} 
  \end{tabular}
  \caption{\label{fig:structure}
    Top view (left) and side view (right) of truncated FeP molecule
    on Au(111) substrate probed by STM tip. First the full FeTPPS molecule
    on the Au(111) surface was relaxed using the VASP code. The four
    phenyl sulfonyl hydroxide groups were then replaced by hydrogen atoms.
    See text for further explanations.
  }
\end{figure}

Our starting point is the structure of the full FeTTPS molecule on the Au(111) surface,
which was relaxed in previous work using the VASP code with the PBE functional in connection
with the van der Waals correction due to Grimme~\cite{Karan:NL:2018}.
As can be seen in Fig.~\ref{fig:structure} (which shows the truncated FeP molecule, see caption
and the discussion below), the porphyrin ring is strongly distorted from a planar geometry due to the binding of the
phenyl sulfonate end groups to the Au substrate, similar to the case of manganese
tetraphenylporphyrin sulfonate (MnTPPS) on Au(111)~\cite{Karan:PRL:2015}.
In the next step we perform ab initio DFT calculations as described above, on a
simplified system consisting of an iron porphyrin (FeP) molecule on the Au(111) substrate and an STM
made of a small [111] pyramid of Au atoms 10~\r{A} above the surface, as depicted in
Fig.~\ref{fig:structure}.
The structure of the FeP molecule is that of the larger FeTPPS molecule on Au(111),
but with the phenyl sulfonate groups substituted by hydrogen atoms.
Hence the local environment of the Fe atom, which determines the crystal field splittings
of the $3d$-orbitals of the Fe center and their hybridization with the substrate
and porphyrin ring, is the same as that of the full FeTPPS molecule.

Previous spin-polarized DFT calculations indicate that the molecule is in a spin-1 state~\cite{Karan:NL:2018}.
The molecular spin $S=1$ is essentially
localized in the $3d$-shell of the Fe center, namely in the $z^2$- and $d_\pi$-orbitals.
The $xy$-orbital is completely full and thus does not carry a spin. The spin in the half-filled
$x^2-y^2$-orbital, on the other hand, is completely quenched by strong coupling to the porphyrin ring,
similar to the case of manganese porphyrin (MnP) on Au(111)~\cite{Karan:PRL:2015}.

Spin-polarized DFT calculations predict the ground state of the FeP molecule to be dominated by a
single electronic configuration where the symmetry of the two $d_\pi$-orbitals is completely broken
due to the saddle-like distortion of the porphyrin ring. The spin-1 is thus almost entirely carried
by the $z^2$- and $yz$-orbitals, while the $xz$-orbital is essentially full~\cite{Rubio-Verdu:NCommP:2018}.
However, it is well known that this static mean-field picture is not really correct for FeP in the
gas phase. Instead a dynamic picture where several electronic configurations close-by in energy,
characterized by different occupations of the Fe $3d$-shell, contribute significantly to the ground
state of the molecule, is more appropriate. Such a dynamic picture of the ground state of FeP derivatives is for
example of fundamental importance for understanding the biologically important imbalance between
carbon dioxide and carbon monoxide binding to the heme protein~\cite{Weber:PNAS:2014}.

\begin{figure}
  \begin{center}
    \includegraphics[width=0.99\linewidth]{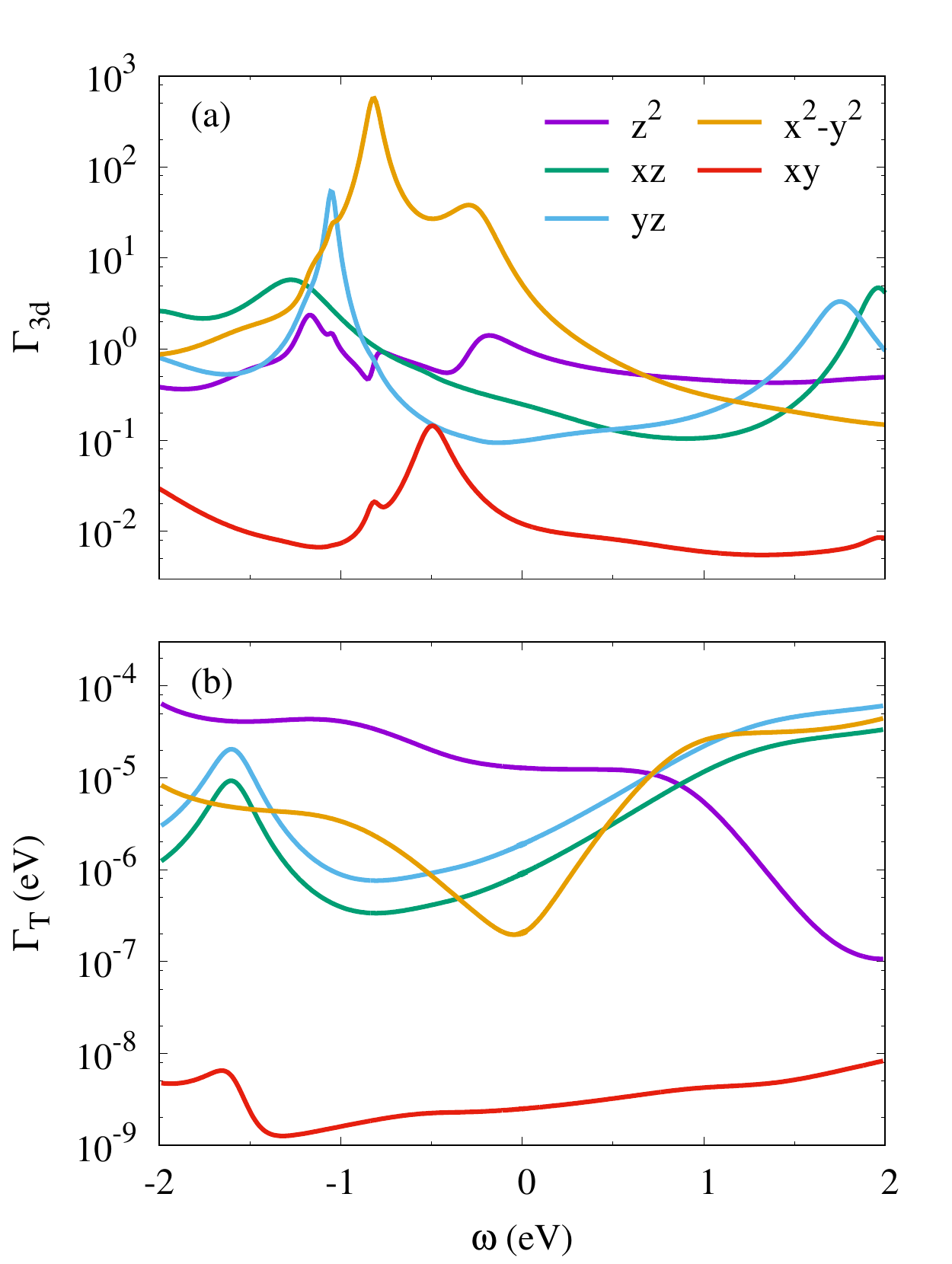}
  \end{center}
  \caption{
    \label{fig:broadening}
    Broadening of Fe $3d$-orbitals of FeP molecule on Au(111) due to coupling to (a) the rest
    of the molecule and surface ($\Gamma_{3d}$) and  (b) to the STM tip placed at 10\r{A} above
    the substrate ($\Gamma_\tip$).
    $\Gamma_{3d}$ is given by the imaginary  part of the hybridization function,
    $\Gamma_{3d}=-2\Im\Delta_{3d}$, while $\Gamma_\tip$ is given by the imaginary part
    the embedding self-energy for the tip electrode, $\Gamma_\tip=-2\Im\Sigma_\tip$,
    which can be obtained from eq.~(\ref{eq:Sigma_tip}).
  }
\end{figure}

As described above in Sec.~\ref{ssec:DFT+AIM} here we make use of a dynamic treatment
of the Fe $3d$-shell in terms of an AIM which is solved within the
one-crossing approximation. This approach takes into account dynamic fluctuations
between different electronic configurations of the $3d$-shell. Some of the model parameters
of the AIM (impurity level energies and hybridization function) are obtained from the DFT
calculation.
In order to not artificially spin-polarize the porphyrin ring and metal substrate, we start
from a \emph{non-magnetic} DFT calculation. Here we have employed the generalized gradient
approximation (GGA) in the parametrization due to Perdew, Burke and Ernzerhof~\cite{Perdew:PRL:96}
in combination with the LanL2MB basis set plus pseudo potential including the outer core and valence electrons~\cite{Hay:JCP:1985}.
Fig.~\ref{fig:broadening}(a) shows the imaginary part of the hybridization function, describing the
broadening of the $3d$-orbitals by the coupling to the porphyrin ring and the Au substrate,
in an energy window of $\pm2$eV around the Fermi level, calculated from the non-magnetic DFT solution
via eq.~(\ref{eq:hybfunc}).

\begin{table}
  \begin{tabular*}{\linewidth}{l@{\extracolsep{\fill}}ccccc}
    Calc.    & $z^2$ & $xz$ & $yz$ & $x^2-y^2$ & $xy$ \\
    \hline
    SP-GGA ($\su$)   & 0.96 & 0.98 & 0.98 & 0.61 & 1.00 \\
    SP-GGA ($\sd$)   & 0.77 & 0.21 & 0.23 & 0.52 & 0.99 \\
    NM-GGA           & 1.50 & 1.49 & 1.34 & 1.06 & 1.99 \\
    OCA ($-0.5$eV) \hspace{1ex} & 1.92 & 1.14 & 1.19 &  --  &  --  \\
    OCA ($+0.0$eV) & 1.90 & 1.11 & 1.15 &  --  &  --  \\
    OCA ($+0.5$eV) & 1.86 & 1.09 & 1.13 &  --  &  --  \\
  \end{tabular*}
  \caption{\label{tab:occupancies}
    Orbital-resolved occupation numbers of Fe $3d$-shell for
    spin-polarized GGA (SP-GGA, occupancies per spin),
    non-magnetic GGA (NM-GGA),
    and OCA calculations of the 3AIM for different shifts $\delta\epsilon$
    (given in parentheses) of the impurity energy levels.
  }
\end{table}

Table~\ref{tab:occupancies} shows the occupancies of the Fe $3d$-orbitals for a spin-polarized
as well as a non-magnetic DFT calculation. The spin-polarized calculation yields a spin-1
basically localized in the half-filled $d_\pi$-orbitals with a small contribution of the $z^2$-orbital.
On the other hand, the non-magnetic solution yields an intermediate valence picture for the three orbitals
$z^2$, $xz$ and $yz$, indicating fluctuations between configurations with different occupations of
these orbitals, which in the spin-polarized DFT calculations is frozen to the lowest energy configuration.
Neither the $x^2-y^2$-orbital nor the $xy$-orbital carries a net spin in the spin-polarized
DFT calculation. While the latter orbital is basically completely full, the former is actually half-filled.
However, the spin in the $x^2-y^2$-orbital is completely quenched due to the strong coupling of
this orbital to the porphyrin ring, indicated by the two strong resonances in the broadening [yellow
 line in Fig.~\ref{fig:broadening}(a)] around $\omega\sim-0.5$eV and $\omega\sim-0.8$eV. This orbital has the
strongest hybridization of all Fe $3d$-orbitals (note the logarithmic scale).
In contrast the fully occupied $xy$-orbital has the weakest hybridization of all $3d$-orbitals, as it
basically does couple to the substrate or porphyrin ring.

Hence it is clear that the spin is localized in the $z^2$- and $d_\pi$-orbitals of the Fe center,
and that fluctuations between different electronic configurations of the FeP molecule chiefly
concern these three orbitals. This justifies the use of a three-orbital AIM to model the dynamics
of the Fe center coupled to the porphyrin ring and Au surface, consisting of the $z^2$-, $xz$- and
$yz$-orbitals only. The broadening of the $z^2$-orbital has a relatively smooth energy dependence around
the Fermi level as it mainly couples to the $s$-type conduction electrons of the Au substrate. 
On the other hand, for symmetry reasons the $xz$- and $yz$-orbitals do not couple to the $s$-type conduction
electrons in the substrate at all, but show significant coupling to molecular orbitals of the porphyrin ring
indicated by the pronounced resonances in the hybridization function of these two orbitals.

Figure~\ref{fig:broadening}(b) shows the broadening of the Fe $3d$-orbitals due to the coupling to the STM tip
10~\r{A} above the surface, computed from the non-magnetic DFT solution via eq.~(\ref{eq:Sigma_tip}).
Comparison with Fig.~\ref{fig:broadening}(a) shows that the tip coupling is smaller by several orders of
magnitude than the coupling to the substrate and the porphyrin ring, thus justifying the assumption $\Gamma_\tip\ll\Gamma_\sub$
made in deriving (\ref{eq:stm_didv}). Importantly, the tip broadening around the Fermi level for the $z^2$-orbital
is larger than that of the other orbitals by at least one order of magnitude, as only this orbital couples
to the $s$-type conduction electrons in the tip due to symmetry reasons. 

\begin{figure*}
  \begin{center}
    \includegraphics[width=0.99\linewidth]{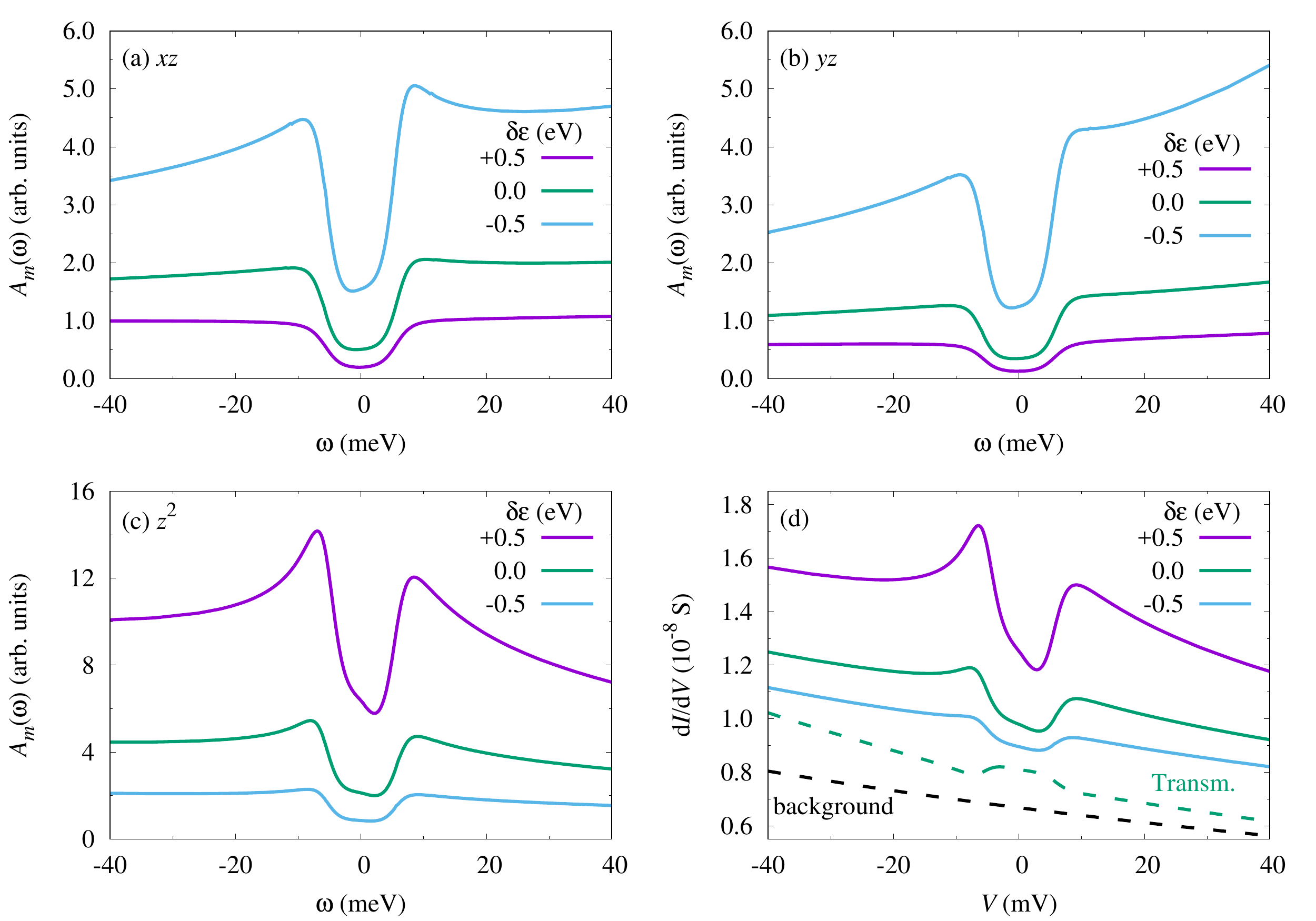}
  \end{center}
  \caption{
    \label{fig:FeP_spectra}
    Calculated spectra for FeP molecule on Au(111) at $T\sim10$K.
    (a-c) Orbital resolved spectral function $A_m(\omega)$ for three of the Fe $3d$-orbitals
    for different shifts $\delta\epsilon$ of the Fe $3d$-level energies $\epsilon_i=h^{3d}_{ii}$
    w.r.t. the value given by the FLL DCC.
    (d) $dI/dV$ spectra computed from eq.~(\ref{eq:stm_didv}) for the geometry depicted in
    Fig.~\ref{fig:structure} with the STM tip 10\r{A} above the substrate for different
    shifts $\delta\epsilon$ of the $3d$-level energies. Impurity parameters: $U=5.3$eV, $U^\prime=3.9$eV, $J_H=0.7$eV, $D=10$meV.
  }
\end{figure*}

Next the 3-orbital AIM consisting of the $z^2$-, $xz$- and $yz$-orbitals
coupled to the Au substrate and porphyrin ring is solved using the one-crossing
approximation, for the hybridization function, shown in Fig.~\ref{fig:broadening}(a),
and impurity levels extracted from the non-magnetic DFT calculation as described
above. For the interaction parameters we use $U=5.3$eV for the intra-orbital Coulomb
repulsion ($U_{iiii}$), $U^\prime=3.9$eV for the inter-orbital Coulomb repulsion ($U_{ijij}$ with $i\neq{j}$)
and $J_H=0.7$eV for the Hund's rule coupling ($U_{ijji}$ with $i\neq{j}$). The uni-axial magnetic anisotropy is
set to $D=10$meV. We have checked that changing the parameters within reasonable bounds
does not alter the results in an essential way.

In Figs.~\ref{fig:FeP_spectra}(a-c) the many-body spectral functions of the impurity orbitals
calculated within OCA at low temperature ($T\sim10$K) are shown. As the DCC is not exactly
known (see above) we have also explored the effect of shifting the impurity levels in energy
by $\pm0.5$eV w.r.t. the value obtained using the FLL DCC. The spectra of all three orbitals
show the typical step features associated with spin-flip excitations of a spin-1 quantum magnet.
It is interesting that even though the spin $S\approx1$ is mainly localized in the half-filled
$xz$- and $yz$-orbitals, with only a minor contribution coming from the nearly full $z^2$-orbital
(see Tab.~\ref{tab:occupancies}), the spin flip excitation steps are clearly visible in the spectra
of all three orbitals. 

The excitation energies are renormalized to about $\tilde\Delta_\pm\sim\pm8$meV w.r.t. the
bare value of $\Delta_\pm=\pm{D}=\pm10$meV due to exchange coupling to the conduction electrons
in the Au substrate and porphyrin ring~\cite{Oberg:NNano:2014,Jacob:EPJB:2016,Jacob:PRB:2018}.
The value for the uni-axial MA of $D=10$meV has of course been chosen such as to reproduce the
experimentally observed spin flip excitation energies for the FeTPPS molecule on the Au(111)
substrate~\cite{Karan:NL:2018}. This value is also in good agreement with the ones measured for
similar iron porphyrin systems on the same substrate~\cite{Li:SciAdv:2018,Rubio-Verdu:NCommP:2018}.

As the impurity level energies are shifted, the spectra of the three orbitals behave quite
differently: While the amplitude of the spin excitations of the $xz$- and $yz$-orbitals
increases when the impurity levels are lowered in energy, the $z^2$-orbital shows just the
opposite behavior. This can be understood by considering the charge fluctuations (measured as the
deviation from integer occupancy) of individual orbitals as a function of the energy shift
(see Tab.~\ref{tab:occupancies}): While the charge fluctuations of the nearly half-filled (singly occupied)
$xz$- and $yz$-orbitals increase, the charge fluctuations of the nearly filled (doubly occupied)
$z^2$-orbital decrease, when the impurity levels are lowered in energy. As was shown in previous
work~\cite{Jacob:PRB:2018}, the amplitudes of the spectral features associated with the spin-flip
excitations increase (decrease) as the charge fluctuations in the corresponding orbitals
increase (decrease). Simultaneously, the asymmetry of the spectra increases (decreases) with
increasing (decreasing) charge fluctuations.

Finally, Fig.~\ref{fig:FeP_spectra}(d) shows the $\rmd{I}/\rmd{V}$-spectra calculated from the
spectral function of the C region according to (\ref{eq:stm_didv}). The STM spectra are dominated
by the spectral function of the $z^2$-orbital plus a linear background stemming from tunneling through
the rest of the molecule.\footnote{The molecular background of the $\rmd{I}/\rmd{V}$ spectrum was
  calculated by shifting the impurity orbitals to very high energies
  ($\sim1000$eV) in order to suppress tunneling into these orbitals.}
The spectral functions of the $xz$- and $yz$-orbitals only contribute indirectly (via coupling to
orbitals on neighboring atoms in the molecule) to the $dI/dV$-spectra, 
since their coupling to the STM tip is suppressed by one order of magnitude compared to that of the
$z^2$-orbital, as discussed above.
The resemblance of the calculated $\rmd{I}/\rmd{V}$ spectrum with the experimentally measured ones
for different iron porphyrin type molecules on the Au surface is quite remarkable~\cite{Karan:NL:2018,Li:SciAdv:2018,Rubio-Verdu:NCommP:2018}.
In contrast the $\didv$ calculated in the phase coherent approximation~\cite{Jacob:PRL:2009} from the correlated
transmission function, $\didv\sim{T(eV)}$ with $T(\omega)=\Tr[\bm\Gamma_\tip\bm{G}^\dagger_\cen(\omega)\bm\Gamma_\sub(\omega)\bm{G}_\cen(\omega)]$,
does not show the typical step features at the excitation energies but rather the inverse behavior with
a smaller amplitude [dashed green line in Fig.~\ref{fig:FeP_spectra}(d)].

The theoretical results suggest that the relatively small differences in the experimental spectra of different
FeP type molecules~\cite{Karan:NL:2018,Li:SciAdv:2018,Rubio-Verdu:NCommP:2018} may to some degree be attributed to small
variations in the occupancy of the Fe $3d$-orbitals, which in turn lead to variations in the spectral functions
according to Fig.~\ref{fig:FeP_spectra}.
The main difference between the experimental and theoretical spectra calculated in the ideal STM limit from eq.~\ref{eq:stm_didv}
seems to lie in the background dispersion which may be attributed to the use of an incomplete basis set and approximate functionals
in the DFT calculations. However, also quantum interference effects in the tunneling between the tip and molecule, not taken into
account in (\ref{eq:stm_didv}), may play a minor role in determining the actual lineshapes of the spin flip excitations in the STM spectra.

\subsection{Kondo effect of Mn porphyrin on Au(111)}

Naturally, also the Kondo effect, which can be seen as a zero-energy or elastic spin flip excitation, %%leading to a Kondo resonance in the spectral function,
can be described within this approach. 
Let us thus briefly revisit the case of a Mn tetraphenylporphyrin sulfonate (MnTPPS) molecule on the Au(111) substrate.
There the STM spectra showed a zero-bias anomaly, which was explained as due to an underscreened Kondo effect in the
$z^2$-orbital strongly enhanced by charge fluctuations~\cite{Karan:PRL:2015}.

\begin{figure}
  \begin{center}
    \includegraphics[width=0.99\linewidth]{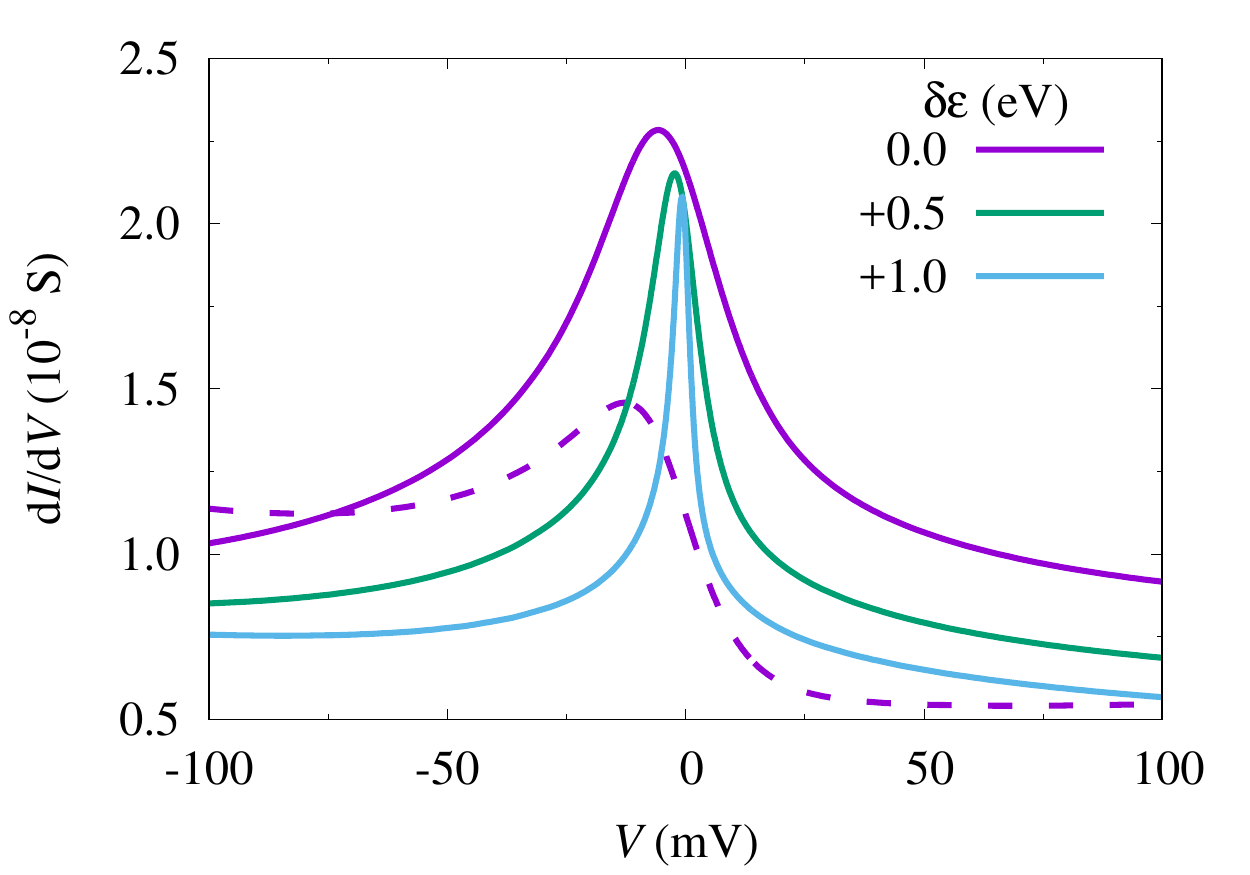}
  \end{center}
  \caption{
    \label{fig:MnP_spectra}
    Calculated $\rmd{I}/\rm{d}V$ spectra for MnP molecule on Au(111) with the STM tip at 10\r{A}
    above the substrate for different shifts $\delta\epsilon$ of the $3d$-level energies.
    The dashed line shows the $\didv$ calculated in the phase coherent approximation
    from the correlated transmission function, $\didv=(2e^2/h)T(eV)$, for $\delta\epsilon=0$.
    Impurity parameters: $U=5.3$eV, $U^\prime=3.9$eV, $J_H=0.7$eV, $D=0$meV.
  }
\end{figure}

Here we calculate the $\didv$ spectrum
using (\ref{eq:stm_didv}) for the truncated MnP molecule on Au(111), also considered in our previous work. Similar to the
truncated FeP molecule considered above, first the full MnTPPS molecule on the Au(111) substrate was relaxed with the VASP
code (see Supplemental Material of Ref.~\cite{Karan:PRL:2015} for more details). Then ab initio DFT calculations using the
PBE functional plus Lanl2MB basis set plus pseudo potentials are performed on a truncated molecule with the four phenyl
sulfonate groups replaced by hydrogen atoms. The Anderson model consisting of the Mn $3d$-shell coupled to the rest of the
system (porphyrin ring plus Au substrate) is solved within the OCA which yields the spectral functions and many-body self-energies
of the Mn $3d$-orbitals.

Finally, the $\didv$ spectra are calculated according to (\ref{eq:stm_didv}) for the STM tip
at 10\r{A} above the surface. The results are shown in Fig.~\ref{fig:MnP_spectra} for different shifts $\delta\epsilon$
of the impurity level energies w.r.t. the value given by the FLL DCC. The $\didv$ shows a Kondo resonance close to zero bias,
stemming from the underscreened Kondo effect in the Mn $z^2$-orbital.
For the impurity level energies given exactly by the FLL DCC, the charge fluctuations lead to a relatively broad and somewhat
asymmetric Kondo resonance. As the impurity levels are shifted upwards in energy the charge fluctuations decrease and lead to a
sharper and more symmetric Kondo resonance.

However, the lineshapes in the experimental STM spectra are considerably more asymmetric than the ones calculated in the
ideal STM limit (\ref{eq:stm_didv}). In fact, the STM spectra calculated using the phase coherent approximation in our
previous work showed a far better agreement with the experimental spectra~\cite{Karan:PRL:2015}.
For comparison Fig.~\ref{fig:MnP_spectra}, also shows the $\didv$ calculated in the phase coherent approximation
from the correlated transmission function $T(\omega)$~\cite{Jacob:PRL:2009,Jacob:JPCM:2015} which shows rather a Fano-like
behavior, not captured by the $\didv$ calculated in the ideal STM limit. Apparently, despite the weak coupling to the STM tip
and in contrast to the inelastic spin flip excitations, quantum interference in the tunneling processes thus still plays a significant
role for determining the actual lineshape of the zero-bias anomaly associated with the Kondo effect. 

\section{Conclusions}
\label{sec:Conclusions}

In this work a semi ab initio scheme for calculating the STM spectra of magnetic molecules
on metallic substrates has been devised, which takes properly into account many-body effects, leading to inelastic
spin flip excitations. The calculated STM spectra of an FeP molecule on the Au(111) substrate show 
the typical step features characteristic for inelastic spin flip excitations, and are
in good agreement with STM spectra recently measured in several
experiments on similar FeP type molecule on Au(111)~\cite{Karan:NL:2018,Li:SciAdv:2018,Rubio-Verdu:NCommP:2018}.
While part of the mild discrepancies between experimental and theoretical spectra may be attributed
to the use of approximate functionals and truncated basis sets in the density functional calculations,
also quantum interference effects, explicitly neglected in the ideal STM limit considered here,
may play a minor role in the end.
In contrast, in the case of Kondo effect quantum interference plays a crucial role
in determining the actual lineshape of the zero-bias anomaly in the STM spectra even for very weak
coupling to the STM tip.
To reconcile the ideal STM limit, properly taking into account inelastic many-body scattering,
with the phase coherent approximation, correctly describing quantum interference,
is the subject of ongoing research.

\begin{acknowledgments}

I am grateful for fruitful discussions with N. Lorente, A. Droghetti, R. Calvo and S. Kurth,
and acknowledge financial support by “Grupos Consolidados UPV/EHU del Gobierno Vasco” (IT578-13).

\end{acknowledgments}

\bibliography{nanodmft}

\end{document}